\documentclass[a4paper,11pt]{article}

\usepackage{pub} 

\usepackage[T1]{fontenc} 
\usepackage{amssymb}
\usepackage{amsmath}
\usepackage{nicefrac}
\usepackage{nccmath}
\usepackage{amsfonts}                   
\usepackage{mathtools}
\usepackage{bbold}
\usepackage{marvosym}
\usepackage{physics}
\usepackage{listings}
\usepackage{relsize}
\usepackage{gauss}
\usepackage{empheq}
\usepackage{cancel}

\usepackage{enumitem}
\usepackage{pdfpages}
\usepackage{graphics}
\usepackage{graphicx}
\usepackage{color}  
\usepackage{soul}  
\usepackage{rotating}
\usepackage{svg}
\usepackage{tikz}
\usepackage{xcolor}
\usepackage{pstricks}
\usepackage{transparent}
\usepackage{soul}
\usepackage[english]{babel}
\usepackage{scrextend}
\usepackage{multirow}
\usepackage[draft]{todonotes}
\usepackage{dirtytalk}
\usepackage[style=english]{csquotes}
\usepackage{tocloft}
\usepackage{ocg-p}
\usepackage{ocgx}
\usepackage{floatrow}
\usepackage{sidecap}
\usepackage{comment}
\usepackage[numbers]{natbib}
\usepackage{epigraph}

\usepackage[bottom]{footmisc}
\usepackage{slashed}
\usepackage{multicol}
\usepackage{fancyhdr}
\usepackage{verbatim}
\newcounter{eg}                         \newtheorem{eg}{Example}[section]        
\def\beg{\begin{eg}\rm}                 \def\eeg{\hfill\sq\end{eg}}

\makeatletter
\DeclareRobustCommand{\cev}[1]{%
  {\mathpalette\do@cev{#1}}%
}
\newcommand{\do@cev}[2]{%
  \vbox{\offinterlineskip
    \sbox\z@{$\m@th#1 x$}%
    \ialign{##\cr
      \hidewidth\reflectbox{$\m@th#1\vec{}\mkern4mu$}\hidewidth\cr
      \noalign{\kern-\ht\z@}
      $\m@th#1#2$\cr
    }%
  }%
}
\makeatother

\def\lr#1{\left(#1\right)}

\usepackage{url}
\usepackage{xurl}
\PassOptionsToPackage{hyphens}{url}
\usepackage{hyperref}
\hypersetup{
    colorlinks=true,
    citecolor=[rgb]{0,0.334,0.666}, 
    linkcolor=[rgb]{0,0.334,0.666},
    urlcolor=[rgb]{0,0.334,0.666},
}

\title{\boldmath Comment on “Geometry of the Grosse-Wulkenhaar model”}

\author[a]{D. Prekrat,
}

\affiliation[a]{Department of Physics and Mathematics, University of Belgrade -- Faculty of Pharmacy,\\ Vojvode Stepe 450, Belgrade, Serbia}

\emailAdd{dragan.prekrat@pharmacy.bg.ac.rs}

\abstract{
We clarify a key point in the geometric reinterpretation of the Grosse-Wulkenhaar (GW) model proposed in “Geometry of the Grosse-Wulkenhaar model” [JHEP 03 (2010) 053]. Specifically, we show that the analysis in Section 6 was performed not for the actual $\Omega$-term in the GW action, which involves both ordinary and star-products, but for a closely related term containing only star-products. Once corrected, the main conclusion—relating the harmonic potential term to background curvature—remains valid, though the parameter identification must be revised. This also resolves a discrepancy concerning the emergence of certain vacuum solutions in the self-dual limit of the model.
}

\keywords{noncommutative QFT, curved spacetime, renormalizability}

\begin{document} 
\maketitle
\flushbottom

\section{Introduction}
\label{intro}
The paper “Geometry of the Grosse-Wulkenhaar model” by Buri\'{c} and Wohlgenannt \cite{Buric:2009ss} explores the differential geometry of a family of curved noncommutative spaces obtained via finite matrix truncation of the Heisenberg algebra. In Section 6, it is claimed that the (two-dimensional) Grosse-Wulkenhaar (GW) model \cite{Grosse:2003nw}
\begin{equation}
S = \int \!\! dx^2 \,
\Bigg(
\frac{1}{2} \partial^\mu\phi \star \partial_\mu\phi
\;+\; \frac{m^2}{2}\phi\star\phi
\;+\; \frac{\Omega^2}{2}(\widetilde{x}^\mu\phi) \star (\widetilde{x}_\mu\phi)
\;+\; \frac{\lambda}{4!}\phi\star\phi\star\phi\star\phi
\Bigg) ,
\label{GW model}
\end{equation}
where the noncommutative $\star$-product encodes noncommutativity of coordinates
\begin{equation}\label{comm-x}
    f \star g
    = 
    f\,e^{\nicefrac{i}{2}\,\cev{\partial}_\mu\Theta^{\mu\nu}\vec{\partial}_\nu}\,g \ ,
    \qquad\qquad
    \comm{x^\mu}{x^\nu}_\star = i\Theta^{\mu\nu} = i\theta \epsilon^{\mu\nu} ,
\end{equation}
and $\widetilde{x}_\mu$ is defined as \cite{Grosse:2004yu}
\begin{equation}\label{tilde-x}
    \widetilde{x}_\mu 
    = 
    2(\Theta^{-1})_{\mu\nu}x^\nu 
    =
    -\frac{2}{\theta}\epsilon_{\mu\nu}x^\nu,
\end{equation}
can be viewed as a field theory on the truncated Heisenberg space with the same algebraic structure but different geometry than the Moyal plane. In this interpretation, the $\Omega$-term responsible for renormalizability appears as a coupling between the scalar field and a background curvature.

However, Section 6 contains several technical inaccuracies. We provide corrections that preserve the core geometric insight but modify the mapping between parameters in the two models.

This study is motivated by a more detailed investigation of the circumstances under which the GW model admits the vacuum solution
\begin{equation}\label{special}
v \star v = \frac{\abs{m^2} - \widetilde{x}^2}{\lambda/6} \ , \qquad m^2 < 0 \ ,
\end{equation}
originally identified in the self-dual case ${\Omega = 1}$ in \cite{deGoursac:2007uv}. Unexpectedly, the matrix formulation of the reinterpreted GW model in \cite{Prekrat:2021uos} admits no such special parameter choice: the same solution appears only when the kinetic term is removed by hand. This discrepancy prompted a re-examination of \cite[Sec. 6]{Buric:2009ss}, which underpins the matrix model construction in \cite{Prekrat:2021uos}.

After including quantum corrections, the solution $v$ may correspond either to the translation symmetry-breaking stripe phase associated with UV/IR mixing or to an asymmetric one-cut phase, depending on the energy. The existence of such an asymmetric one-cut phase---tentatively supported by Monte Carlo simulations \cite{Prekrat:2023thesis}---provides a heuristic lower bound on the triple-point shift in the GW model \cite{Prekrat:2022sir}, which apparently underlies the model's renormalizability \cite{Prekrat:2021uos, Prekrat:2023fts}.

\section{$\Omega$-term reinterpretation}

Let us first note that \eqref{GW model} is written in the $\star$-product formalism, whereas \cite{Buric:2009ss} employs the frame formalism. In this section we compare the results obtained in the two formulations by switching between them; for clarity, we denote the field by $\phi$ in the former and by $\varphi$ in the latter.

Section 6 of \cite{Buric:2009ss} reinterprets the $\Omega$-term as a coupling between the field and the curvature of the truncated Heisenberg algebra. It begins by connecting the kinetic and $\Omega$-terms via
\begin{equation}\label{term mixing v1}
    \int \widetilde{x}^\mu \varphi \widetilde{x}_\mu \varphi =
    \int -\frac{1}{2} \comm{p^\mu}{\varphi} \comm{p_\mu}{\varphi} 
    + \widetilde{x}^\mu \widetilde{x}_\mu \varphi^2 .
\end{equation}
This uses the identification of $\widetilde{x}_\mu$ as $i p_\mu$, which is incorrect. Equation\footnote{
Throughout the text, we adopt the shorthand notation ([$r$].$s$.$n$) to refer to “equation ($s$.$n$) from reference [$r$]”.} (\cite{Buric:2009ss}.5.1) actually states:
\begin{equation}
    \epsilon p_1 = +i\mu^2 y\ ,
    \qquad
    \epsilon p_2 = -i\mu^2 x \ .
\end{equation}
By comparing the commutation relations (\cite{Buric:2009ss}.3.12) and \eqref{comm-x} to identify $\epsilon = \theta \mu^2$, and using \eqref{tilde-x} to express $x^\mu$ in terms of $\widetilde{x}^\mu$, one obtains
\begin{equation}
    p_1 = -\frac{i}{2} \widetilde{x} \ ,
    \qquad
    p_2 = -\frac{i}{2} \widetilde{y} \ ,
\end{equation}
hence $\widetilde{x}_\mu = 2i p_\mu$. In addition, $\widetilde{x}^\mu\widetilde{x}_\mu$ equals ${+ 4\mu^4 x^\mu x_\mu / \epsilon^2}$, not ${- \mu^4 x^\mu x_\mu  / \epsilon^2}$
as claimed in \cite{Buric:2009ss}. This changes \eqref{term mixing v1} into
\begin{equation}\label{term mixing v2}
    \int \widetilde{x}^\mu  \varphi \widetilde{x}_\mu \varphi
    =
    \int \frac{1}{2} \comm{\widetilde{x}^\mu}{\varphi} \comm{\widetilde{x}_\mu}{\varphi} 
    + \widetilde{x}^\mu \widetilde{x}_\mu \varphi^2
    =
    \int -2 \comm{p^\mu}{\varphi} \comm{p_\mu}{\varphi} 
    + \frac{4\mu^4}{\epsilon^2} x^\mu x_\mu \varphi^2.
\end{equation}

Note, however, that term ${\widetilde{x}^\mu \varphi \widetilde{x}_\mu \varphi}$ does not actually appear in the GW action \eqref{GW model}. Namely, since the algebra product used in \cite{Buric:2009ss} maps to the $\star$-product, this term corresponds to ${\widetilde{x}^\mu \star \phi \star \widetilde{x}_\mu \star \phi}$ instead of  ${(\widetilde{x}^\mu \phi) \star (\widetilde{x}_\mu \phi)}$. Let us connect these two terms:
\begin{align}
    \int \widetilde{x}^\mu \star \phi \star \widetilde{x}_\mu \star \phi
    &= \notag
    \int (\widetilde{x}^\mu \star \phi) (\widetilde{x}_\mu \star \phi)
    =
    \int (\widetilde{x}^\mu \phi + i \partial^\mu \phi) (\widetilde{x}_\mu \phi + i \partial_\mu \phi)
    \\
    &= \notag
    \int 
    - \partial^\mu \phi \, \partial_\mu \phi + \widetilde{x}^\mu \phi \, \widetilde{x}_\mu \phi
    + 2i \widetilde{x}^\mu \phi \partial_\mu \phi
    \\
    &= 
    \int 
    - (\partial^\mu \phi) \star (\partial_\mu \phi) + (\widetilde{x}^\mu \phi) \star (\widetilde{x}_\mu \phi)
    + 2i \widetilde{x}^\mu \phi \partial_\mu \phi \ ,
\end{align}
where we utilized an easy-to-prove identity \cite{deGoursac:2007uv}
\begin{equation}
    \widetilde{x}^\mu \star \phi = \widetilde{x}^\mu \phi + i \partial^\mu \phi \ .
\end{equation}
The last term in the integral vanishes due to $\epsilon$-$\delta$ contraction since
\begin{equation}
    \int \widetilde{x}^\mu \phi \partial_\mu \phi 
    = 
    \frac{1}{2} \int \widetilde{x}^\mu \partial_\mu \phi^2
    =
    - \frac{1}{2} \int \partial_\mu\widetilde{x}^\mu \phi^2
    =
    \frac{\cancelto{0}{\epsilon^{\mu\nu}\delta_{\mu\nu}}}{\theta} \int \phi^2.
\end{equation}
Thus, the $\Omega$-term
\begin{equation}
    \frac{\Omega^2}{2} \int (\widetilde{x}^\mu \phi) \star (\widetilde{x}_\mu \phi)
\end{equation}
can be rewritten as
\begin{equation}
    \frac{\Omega^2}{2} \int (\partial^\mu \phi) \star (\partial_\mu \phi) + 
    \widetilde{x}^\mu \star \phi \star \widetilde{x}_\mu \star \phi \ ,
\end{equation}
or, using the operator version \eqref{term mixing v2}, as
\begin{multline}
    \frac{\Omega^2}{2} \int \comm{p^\mu}{\varphi} \comm{p_\mu}{\varphi} + 
    \lr{-2 \comm{p^\mu}{\varphi} \comm{p_\mu}{\varphi} 
    + \frac{4\mu^4}{\epsilon^2} x^\mu x_\mu \varphi^2}
    = \\ =
    \frac{\Omega^2}{2} \int -\comm{p^\mu}{\varphi} \comm{p_\mu}{\varphi} 
    + \frac{4\mu^4}{\epsilon^2} x^\mu x_\mu \varphi^2 .
\end{multline}
This modifies the kinetic coefficient $\kappa$ in (\cite{Buric:2009ss}.6.2) from $1-\Omega^2/2$ to $1-\Omega^2$. The action becomes:
\begin{equation}
    S = \int 
    \frac{1}{2}\lr{1-\Omega^2} \partial^\mu \varphi \partial_\mu \varphi
    \;+\;
    \frac{m^2}{2} \varphi^2 
    \;+\;    
    \frac{2\Omega^2\mu^4}{\epsilon^2} x^\mu x_\mu \varphi^2
    \;+\;
    \frac{\lambda}{4!}\varphi^4,
\end{equation}
now correctly reflecting the self-duality point ${\Omega = 1}$. 
We can finally express the action in terms of the curvature (\cite{Buric:2009ss}.5.8) 
\begin{equation}
    R = \frac{15\mu^2}{2} - 8\mu^4 (x^2 + y^2)
\end{equation}
as:
\begin{equation}\label{new action}
    S = \int 
    \frac{1}{2}\lr{1-\Omega^2} \partial^\mu \varphi \partial_\mu \varphi
    \;+\;
    \frac{1}{2}
    \lr{m^2 +  \frac{15}{4} \frac{\Omega^2}{\epsilon^2}\mu^2} \varphi^2 
    \;-\;
    \frac{1}{4}\frac{\Omega^2}{\epsilon^2} R \varphi^2
    \;+\;
    \frac{\lambda}{4!}\varphi^4.
\end{equation}

Since the action for scalar field non-minimally coupled to curvature is given in \cite{Buric:2009ss} by
\begin{equation}
    S' = \int \sqrt{g}
    \lr{
    \frac{1}{2}\partial^\mu \varphi \partial_\mu \varphi
    \;+\;
    \frac{M^2}{2}\varphi^2 
    \;-\;
    \frac{1}{2}\xi R \varphi^2
    \;+\;
    \frac{\Lambda}{4!}\varphi^4
    },
\end{equation}
and restricting to the present case where $\sqrt{g} = 1$,
the two actions can be identified up to an overall normalization, $S = \kappa S'$. Equation \eqref{new action} then leads to revised parameter identifications (\cite{Buric:2009ss}.6.5):
\begin{equation}
    \kappa = 1-\Omega^2 \, ,
    \qquad
    \frac{\Omega^2}{\epsilon^2} = 2\kappa\xi \, ,
    \qquad
    m^2 = \kappa \lr{M^2 - \frac{15}{2} \xi \mu^2} ,
    \qquad
    \lambda = \kappa\Lambda \, .
\end{equation}
Equation (\cite{Buric:2009ss}.6.6) is modified accordingly, and the self-duality point ${\Omega = 1}$ is now reached in the limit
\begin{equation}
    \kappa \to 0^+ \, ,
    \qquad
    \xi \to \infty \, ,
    \qquad
    M^2 \to  \pm\infty \, ,
    \qquad
    \Lambda \to \infty \, ,
\end{equation}
where the sign of $M^2$ is determined by the sign of
\begin{equation}
    m^2 + \frac{15}{4\epsilon^2}\mu^2 \, .
\end{equation}

\section{Conclusion and Discussion}

In summary, we clarified a key step in the interpretation of the GW model as a scalar field theory on truncated Heisenberg space with curvature coupling, as proposed in \cite{Buric:2009ss}. Specifically, we show that the analysis in Section 6 was applied not to the actual $\Omega$-term in the GW action, which mixes ordinary and star-products, but to a closely related term built solely from star-products. Once corrected, the main conclusion—relating the harmonic potential term to background curvature—remains valid, though the parameter identification must be revised. 

This correction also resolves the initial inconsistency regarding the special vacuum solution. Namely, the revised parameter identifications propagate into the matrix model formulation given in Appendix A of \cite{Prekrat:2021uos}, where the parameters $c_2$, $c_4$, and $c_r$ scale as $1/\kappa$ and thus diverge as $\kappa \to 0$ near self-duality, while the kinetic term remains unaffected. This contrasts with the previously assumed, but incorrect, $\kappa \to 1/2$ limit. As a result, the non-kinetic terms dominate the dynamics in the $\kappa \to 0$ regime, explaining why the vacuum solution \eqref{special} emerges only when the kinetic term becomes negligible.

Equivalently, multiplying the equation of motion (\cite{Prekrat:2021uos}.14) by $\kappa$ introduces a prefactor $\kappa$ in front of its kinetic part, while the remainder of the equation becomes $\kappa$-independent. In the $\kappa \to 0$ limit, this effectively eliminates the kinetic contribution and ensures the existence of the special solution (\cite{Prekrat:2021uos}.16), which is equivalent to \eqref{special}.

This reconciles the results of \cite{deGoursac:2007uv} and \cite{Prekrat:2021uos}, and importantly, it leaves intact the core arguments in \cite{Prekrat:2021uos} that link the shift of the triple point to the model’s renormalizability.

\acknowledgments

This research was supported by the Ministry of Science, Technological Development and Innovation, Republic of Serbia through grants to the University of Belgrade -- Faculty of Pharmacy No. 451-03-136/2025-03/200161 and No. 451-03-137/2025-03/200161.

\bibliographystyle{JHEP}
\bibliography{bibliography}

\end{document}